\documentclass[sigconf,nonacm]{acmart}

\usepackage{algorithmic}
\usepackage{graphicx}
\usepackage{textcomp}
\usepackage{xcolor}
\usepackage{hyperref}

\usepackage{float}
\usepackage{adjustbox}

\usepackage[caption=false]{subfig}

\setcopyright{acmlicensed}
\copyrightyear{2024}
\acmYear{2024}
\acmDOI{XXXXXXX.XXXXXXX}

\acmConference[Conference acronym 'XX]{Make sure to enter the correct
  conference title from your rights confirmation emai}{June 03--05,
  2018}{Woodstock, NY}
\acmPrice{15.00}
\acmISBN{978-1-4503-XXXX-X/18/06}

\begin{document}

\title{Evaluation of Reinforcement Learning for Autonomous Penetration Testing using A3C, Q-learning and DQN}

\author{Norman Becker}
\affiliation{%
  \institution{German Research Center for Artificial Intelligence (DFKI)}
  \city{Kaiserslautern}
  \country{Germany}}
\email{norman.becker@dfki.de}
\orcid{0009-0008-5575-1393}

\author{Daniel Reti}
\affiliation{%
  \institution{German Research Center for Artificial Intelligence (DFKI)}
  \city{Kaiserslautern}
  \country{Germany}}
\email{daniel.reti@dfki.de}
\orcid{0000-0001-8071-6188}

\author{Evridiki V.Ntagiou}
\affiliation{%
  \institution{OPS-GDA, ESA-ESOC}
  \city{Darmstadt}
  \country{Germany}}
\email{evridiki.ntagiou@esa.int}
\orcid{0000-0003-3403-2863}

\author{Marcus Wallum}
\affiliation{%
  \institution{OPS-GDA, ESA-ESOC}
  \city{Darmstadt}
  \country{Germany}}
\email{Marcus.Wallum@esa.int}
\orcid{0009-0004-3306-856X}

\author{Hans D. Schotten}
\authornote{Also with Department of Electrical and Computer Engineering, Technische Universität Kaiserslautern.}
\affiliation{%
  \institution{German Research Center for Artificial Intelligence (DFKI)}
  \city{Kaiserslautern}
  \country{Germany}}
\email{hans_dieter.schotten@dfki.de}
\orcid{0000-0001-5005-3635}

\renewcommand{\shortauthors}{Becker, et al.}

\begin{abstract}
Penetration testing is the process of searching for security weaknesses by simulating an attack. It is usually performed by experienced professionals, where scanning and attack tools are applied. By automating the execution of such tools, the need for human interaction and decision-making could be reduced. In this work, a Network Attack Simulator (NASim) was used as an environment to train reinforcement learning agents to solve three predefined security scenarios. These scenarios cover techniques of exploitation, post-exploitation and wiretapping. A large hyperparameter grid search was performed to find the best hyperparameter combinations. The algorithms Q-learning, DQN and A3C were used, whereby A3C was able to solve all scenarios and achieve generalization. In addition, A3C could solve these scenarios with fewer actions than the baseline automated penetration testing. Although the training was performed on rather small scenarios and with small state and action spaces for the agents, the results show that a penetration test can successfully be performed by the RL agent.
\end{abstract}



\maketitle


\section{Introduction}
\label{sec:introduction}
Although many steps of a penetration test are automated to a high degree, it still needs experience and skill to correctly operate such tools, avoid detection and have high confidence that no vulnerability was overlooked.
With the ever-evolving technologies and advancing threat landscapes, protecting information systems requires constant testing against the state-of-the-art vulnerabilities and exploitations of these. As this process requires highly skilled and certified personnel, which is scarce, there is a high demand for automating the security testing process. In a penetration test, an attack on a computer network is simulated with the goal of penetrating the perimeters of the network, such as firewalls or Intrusion Detection Systems (IDS), and show that a successful compromise of sensitive hosts is possible. Traditionally, to achieve this, a team of human experts will operate a variety of customly written or off-the-shelf tools to scan the networks, hosts and services and identify and execute vulnerability exploitations, e.g. with Nmap or Metasploit.
In recent years, many publications have looked into automated penetration testing based on reinforcement learning (RL) \cite{Standen.2021,Tran.2021,MicrosoftDefenderResearchTeam.2021,Schwartz.2019}, where an agent can choose from predefined actions to interact with an environment and adjust its behaviour based on the cost and reward of the actions.
Existing approaches of reinforcement learning-based (RL) penetration testing use insufficient metrics to evaluate the performance of their agents as they compare them against random action selection, only considering if the agent can solve an environment or train on a single environment and ignore the ability of generalization, which makes the agent inapplicable for real-world environments. Additionally, certain common attack scenarios and actions are missing, depending on the framework.  

In this work, the authors extended a Network Attack Simulation environment (NASim) for RL developed by Schwartz et al. \cite{Schwartz.2019}, by adding additional actions, defining four attack scenarios, and introducing a baseline agent based on decision trees (Penbox). Permutations of each scenario were created to address the generalisation and to avoid overfitting. The authors split the training of the RL agents into different stages of complexity:
\begin{enumerate}
    \item Stage 1: 24 Permutations of one Scenario 
    \item Stage 2: 72 Permutations of all three Scenario 
    \item Stage 3: Test generalization against unseen permutations
\end{enumerate}
As RL agents, Q-Learning, DQN and A3C were used. 
To find the agent with the best combination of hyperparameters, a large hyperparameter grid search was performed on a total of more than 4000 agents.

A3C solved all three stages, while DQN learning failed in all stages. Q-learning could solve Stage 1 in one scenario. 
The A3C agent required fewer actions on average compared to a baseline decision tree.

The contributions of this paper are as follows:
\begin{itemize}
    \item Extension of NASim \cite{Schwartz.2019} with additional functionalities such as Vulnerability Scanning, Wiretapping and \\ Post-Exploitation.
    \item Specification of three different attack scenarios, each requiring different attack actions for a successful exploit.
    \item An extensive hyperparameter grid search has been performed to identify the best combination of hyperparameters.
    \item Evaluation of Q-Learning, DQN-learning and A3C-based agents for autonomous penetration testing. To the authors' best knowledge, this work is one of the first to apply A3C to automated penetration testing.
\end{itemize}

The artefacts for this paper can be found on GitHub.\footnote{\url{https://anonymous.4open.science/r/Paper\_public-EA2C/}}
The remainder of this paper is structured as follows. Chapter II provides background information on the concept of penetration testing and its automation. In Chapter III, presents related work. In Chapter IV the methodology and experiment design is described, and the results are shown in Chapter V. In Chapter VI, the results are discussed. Lastly, Chapter VII concludes this work.

\section{Penetration Testing}
A first definition for penetration testing was given by R. R. Linde in 1975, where it was described as the process of assessing vulnerabilities in the form of design weaknesses in the implementation of an operating system's security controls. It was pointed out that it is a cost-effective alternative to more formal correctness proof, which is not available \cite{Linde.1975}. 
Today, penetration testing describes the proactive approach of searching for security weaknesses in the design or implementation of a network, device or software by using tools and methods similar to those of a malicious attacker. 

\subsection{Fundamentals}
This process of simulating an attack is sometimes also referred to as red teaming, a military term for the attacking team in military simulations to test defence tactics, or as ethical hacking, in order to distinguish from the malicious attacker, which is often referred to as a 'hacker' in popular culture.

As it is impossible to acquire a complete list of every possible vulnerability to test for, a penetration test can never prove security \cite{Geer.2002}. Therefore, when a penetration test does not find any vulnerabilities, it is not proven that there is no presence. The value of a penetration test comes from the correct interpretation of the results and requires an understanding of the present threats and attacker capabilities.

Due to this reliance on experience and knowledge, a penetration test is usually performed by a security professional. Nonetheless, a lot of automation in penetration testing tools exists to assist the human operator, who is interpreting scanning results, deriving suitable tactics and tools and reporting the results. 

Such a penetration test is only valid for the time it is pursued and needs to be regularly repeated as new vulnerabilities are found, and system configurations may be changed. 

    \subsection{Scope and Phases}
    The goal of every penetration test is to find potential vulnerabilities in a system. Hereby the amount of information provided beforehand determines if it is considered a black box, grey box or white box testing. With black box testing, the information and access to the system for the tester are the same as an outsider attacker. With white box testing, the tester is provided with all information beforehand, similar to the information a malicious insider could have, which allows them to find more sophisticated vulnerabilities which might stay undetected from a black box test.
    The target of the test can be a single piece of software application but also a whole system, where the combination of an operating system, services installed, individual configuration and security controls might pose an attack surface, or it could be the whole network composed of servers, clients, services and networking equipment.
    While there is no standard taxonomy of the phases of a penetration test, it typically it consists of the following phases:
    \begin{itemize}
     \item planning
     \item information gathering or reconnaissance
    \item exploitation attempt
    \item post-exploitation
    \item reporting.
    \end{itemize}
Part of the scope could also be the personnel to which social engineering might be applied.
For this work, the scope is only the information gathering and exploitation phase and post-exploitation, and the mode is black-box testing
    
    \subsection{Automation of Pentesting}
An innumerable amount of free and commercial tools exist to help facilitate and automate the process of security testing. The most commonly known tool suite is Kali Linux and the Metasploit framework. The purpose of such tools could be, for example, scanning, sniffing, brute forcing, password or hash cracking, fuzzing, payload generation, debugging, static analysis, dynamic analysis, delivery and handling of remote sessions. A lot of automation is already achieved by combining the execution of such tools, e.g. OpenVAS, Metaspoilt Framework, nmap scripting engine or Nessus. These automation approaches allow to probe for a broad range of known vulnerabilities by going through predefined test cases and probes. 
It should be noted that there are no universally accepted definitions of automated or autonomous penetration testing, and the terms are often used interchangeably. A penetration test that is not automated is considered a manual test. A manual penetration test can be performed in an exploratory or systematic manner. The automated approach can reduce a lot of time and repetition for the systematic test. A vulnerability scan is not a penetration test, as there is no exploitation.  A reinforcement learning-based agent might improve the selection of test cases further to reduce the number of actions needed. A specific systematic solution is presented in the following, which is used as the baseline for comparison with the reinforcement learning-based approach presented in this work. An extensive overview of reinforcement learning-based approaches can be found in Chapter VI.

    \subsubsection{Penbox}
    Penetration testing in a Box, short Penbox, is an automated penetration testing software developed by
    [redacted for blind review].
    It aims to perform security tests against a network and its systems by attacking them with standard penetration testing tools. The overall goal is to discover potential attack paths and vulnerabilities. The software acts like a toolbox and uses decision trees to execute these tools in a meaningful order to perform attacks successfully. Its implementation makes it possible to generate new scenarios, attack trees or modify tools to apply them to other custom scenarios.
    In this work, Penbox is used to collect real-world data and as a baseline to compare the attack trees against the reinforcement learning methods.




\section{Related Work}
It is foreseeable that artificial intelligence will be able to outperform human penetration testing experts \cite{Ghanem.2018}. In recent years, the research topic of automating penetration testing using reinforcement learning has grown. Most publications either focus on either the development of environments or the agents. \\

 Introduced by J. Schwartz in 2019, NASim was one of the first approaches of providing a simulated environment specialized in performing penetration testing with reinforcement learning \cite{Schwartz.2019}. The work has shown that RL agents can find attack paths and solve environments if the network isn't too large. The same author was involved in a work by Baillie et al., who introduced an environment named CybORG in 2020 \cite{Standen.2021}, where in addition to the simulation of the network an emulation is provided as well. The simulation aspect allows for faster training of the agents, while the emulation can tests the agent on real-world systems. CybORG does not only focus on the attacker's perspective, but considers autonomous cyber operations, including agent development for blue teaming and red teaming. \\
Following NASim and CybORG, Microsoft has published a simulator in 2021 called CyberBattleSim \cite{MicrosoftDefenderResearchTeam.2021} with equal base functionalities and provides additional features such a credential handling, and it uses a slightly different model to represent the network.\\
Another RL training environment for penetration testing is CyGil by Li et al. \cite{Li.2021}, published in 2021. CyGil completely emulates networks and doesn't simulate them. With a sufficiently abstracted gym-like interface, it enables reinforcement learning and provides better performances regarding training time compared to a real-world network. Similarly to CybORG, CyGil also provides a blue team mode. Similarly to this work, the authors of NASimEmu \cite{Janisch.26.05.2023} modified NASim to emulate the networks simulated in NASim. They highlight that the handling of credentials is missing in the NASim, which was addressed and solved in this paper. \\

While gym environments are a standardized method to compare algorithms, Zennaro and Erdodi \cite{Zennaro.26.05.2020} applied Reinforcement Learning on simplified Capture-the-Flag challenges, allowing the agent to perform a port scan or analyze a website. In these scenarios, prior knowledge is provided, and the action space for the agent is kept small respectively. Niculae et al. present a formalised game of penetration testing and show that reinforcement learning can outperform humans in finding the most efficient action sequence \cite{Niculae2018AutomatingPT}. Similarly, Turner et al. defined a zero-sum game to represent the penetration testing task \cite{Turner.2022}.
Especially for the penetration testing domain, the agents must handle a large action and state space, and with that, most agents become unstable. This problem is specific for penetration testing in the RL context \cite{Schwartz.2019}. To solve it, new and more robust RL methods, or strong constraints on the action space are needed. One approach is presented by Hu et al. \cite{Hu.2020}. They first create the full attack tree of each topology and then let the agent select the best possible attack paths. The same techniques were later used by Tyler Code, who created a Layered Reference Model \cite{Cody.14.06.2022}. Masking the action space can also minimise options for the agent \cite{Yang.22.02.2022}. Khuong Tran et al. proposed a promising approach to keep the full action space without any limitations by creating an agent based on multiple sub-agents connected in a hierarchical order \cite{Tran.2021}. The agent uses a tree structure including multiple smaller agents, so each agent has a small action space, but all agents chained can provide a large action space. It is unclear whether one agent can solve such tasks and whether splitting the agent up to perform sub-tasks is necessary.
Other approaches try to increase the performance by combining and extending already existing algorithms \cite{Wang.2022}.\\
Another critical aspect of penetration testing is the generalization. As the paper of Cobbe at al. \cite{Cobbe.2018} highlights, letting an agent transfer knowledge to another environment is difficult for state-of-the-art deep reinforcement learning algorithms. 
Most research focuses on solving single environments with often heavily simplified tasks. The small training environments and action spaces make it impossible for these specialized reinforcement learning agents to perform a complete penetration test. In this work, the authors focus on a larger diversity of attack scenarios and environments to generalise better while keeping the full action space for all scenarios. For the evaluation of the agents, other publications have been testing whether the agents could solve the environment successfully or compare the performance to that of a random agent. This paper is the first to compare the RL agents to a baseline decision tree instead of comparing to a random agent, and emphasizes on evaluating whether the agents are sufficiently generalizing.

\section{Experiment}
    \subsection{Scenarios}
    Three scenarios have been defined as the foundation of the experiments. They will be used to train the reinforcement learning agent and then compare the results against the Penbox execution. As a benchmark, these scenarios are also rebuilt in a physical network with Raspberry Pi's (model 3b and 4). Raspberry Pi's are cheap and small one-board computers which support all relevant functions of a real PC. The authors have chosen physical hardware instead of virtual machines because they are more demonstrative and provide better results on real-world data. However, it would also be possible to use virtual machines. \\
    In all scenarios, one attacker aims to compromise as many machines as possible. A machine is compromised when root access is gained. Every topology has a flat hierarchy and requires different hacking methods to be solved, such as Exploits, Wiretapping, and Post-exploitation. For each of these Scenarios, a perfect attack path is known, which the RL agents can be compared with as a baseline. \\
    
        \subsubsection{Exploits}
\begin{figure}[htbp]
        \centerline{\includegraphics[width=1.\columnwidth]{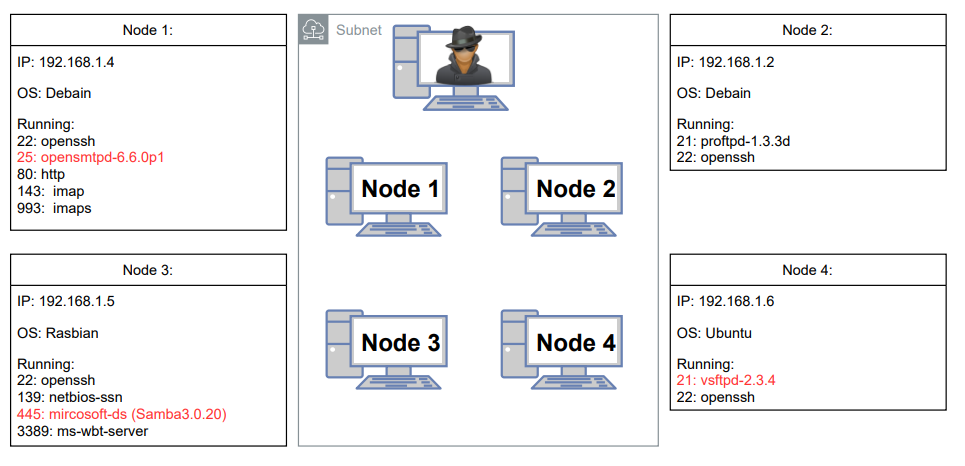}}
        \caption{Scenario A: Exploits. Red-marked services can be exploited to achieve root access.}
        \Description[Showing a Network Topology A ]{The images show a network configuration in which hosts are one subnet together with one attacker. The different open ports are listed for each machine. Important is, that 3 out of the 4 machines got a vulnerable service running, which can be exploited. }
        \label{fig:topologyA}
        \end{figure}

        The first scenario A, with the name Exploits, represents one of the most basic mechanics of penetration testing: exploiting software vulnerabilities in running services. Services are programs running on a system and provide applications like file transfer protocol (FTP), website servers etc. If a vulnerable service is listening to an incoming connection, the service could be used as an entry point to the system. 
        After a software vulnerability is disclosed, it is tracked in vulnerability databases with the \textit{Common Vulnerabilities and Exposures} (CVE) system \cite{TheMITRECorporation.}. 
        Such software vulnerabilities can be exploited by a proof-of-concept exploit script, which can be found in public exploit databases or git repositories \cite{OffensiveSecurity.}. These exploits only work for specified service and software versions. 
        As shown in Figure \ref{fig:topologyA}, scenario A includes four nodes and one attack node in one subnet. Each node represents a system with different services and ports running. The following services are present: SSH, SMTPD, FTP, XRDP and Samba. All nodes have OpenSSH with password authentication running to be accessible with the proper credentials. In this case, the credentials are secure and can not be found by an attacker.
        
        The first node runs opensmtpd-6.6.0p1, which contains the vulnerability \textit{CVE-2020-7247}. If this service runs in a specific configuration, the attacker can execute remote commands, leading to a root shell. 
        Node three is similar to node one. It is running Samba3.0.20 with a vulnerability \textit{CVE-2007-2447}, which allows us to execute arbitrary commands and gain a reverse shell with root permissions if the configuration is fitting.\\
        Node four is running vsftpd-2.3.4. It is exploitable, but not because of a bug or misconfiguration. The service has a malicious backdoor injected due to a supply chain attack on the software's source repository, which made it possible to gain a root shell to the machine by sending the correct string. In light of that, the service is not marked with a CVE.\\
        Node two is running proftpd-1.3.3d, which has no known vulnerability. \\
        Overall, in this scenario, the attacker can compromise three out of four systems by just performing service scans and executing the right exploits.\\
        
        \subsubsection{Wiretapping}
\begin{figure}[htbp]
        \centerline{\includegraphics[width=\columnwidth]{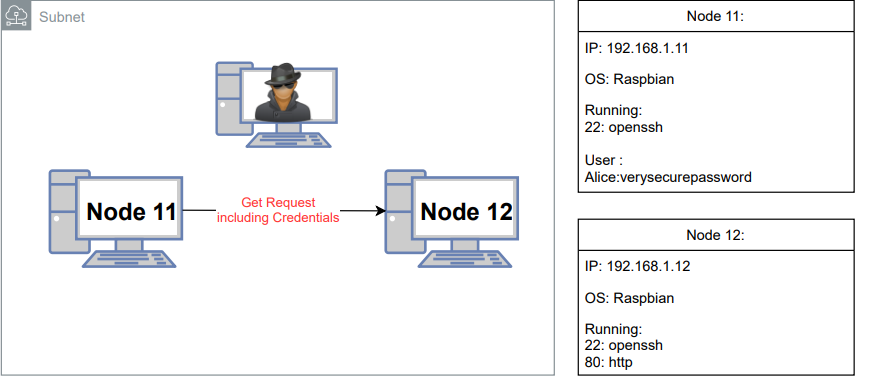}}
        \caption{Scenario B: Wiretapping. Red-marked GET requests transmit credentials in clear text}
        \Description[Showing Network Topology B ]{In the image, a network topology with two hosts can be seen. One Machine is sending a get request to the other machine. The get request contains credentials, which are not encrypted and can be intercepted.}
        \label{fig:topologyB}
        \end{figure}

        The second scenario, Scenario B, represents another penetration testing technique, a specific type of Wiretapping. In case the traffic is unencrypted, it is possible to perform wiretapping to obtain confidential information such as passwords sent over HTTP or FTP. In this scenario \ref{fig:topologyB}, there is no encryption, so sensitive data can be gained without decrypting messages.
        
        Node 12 hosts a standard web application that only authorized users can access. Alice, operating on Node 11, sends a GET request with her credentials attached every 30 seconds to the web application. Due to the fact that Node 12 is using HTTP, every message is sent in clear text. If an attacker can catch such a request packet, he can read the username and the password. If the attacker is inside the network, he can get access to the packets in each of the following cases: 
        \begin{enumerate}
            \item If the subnet is wired, a hub routes the packets instead of a switch.
            \item If the subnet is wired, the attacker could connect to a mirror port, which receives all traffic.
            \item The subnet is wireless, the attack can receive the packages directly and does not need direct physical access.
            \item A proxy can also guide the traffic directly to the attacker. 
        \end{enumerate} 
        If one of the above cases is confirmed, an attacker could use a tool like Wireshark \cite{Wireshark.} to capture all traffic and filter for important information.
        This password is identical to her login password and can then be used to gain access to the machine on Node 11.
        In this scenario, it is possible to compromise one out of two systems. \\
        Unencrypted traffic might not be realistic any more. Nevertheless, it is used as a placeholder for simplicity reasons. Currently, a standard Wireshark filter is used. Encrypted traffic can be cracked by additional steps or information, which can be addressed in feature work. 
        
        \subsubsection{Post-exploitation}
\begin{figure}
        \centerline{\includegraphics[width=\columnwidth]{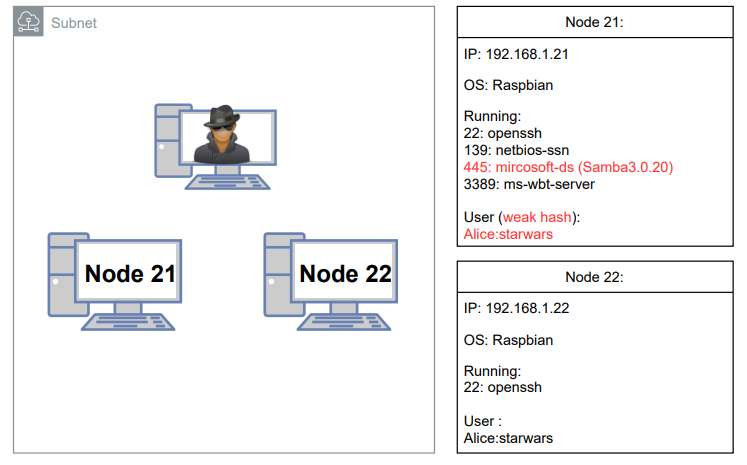}}
        \caption{Scenario C: Post-exploitation. Red-marked services can be exploited; afterwards, credentials are found because of a weak hash.}
        \Description[ToDo]{ToDo}
        \label{fig:topologyC}
        \end{figure}

        The last Scenario C, Post-exploitation, was designed to represent post-exploitation. Post-exploitation is a catch-all name for the steps an attacker performs after gaining access to a system. During this phase, the attacker can try to contact new systems, gather information or install malware to gain persistence. In this scenario, the attacker can gain new user credentials after successfully compromising a system. With this information, he can access another system. \\
        In this scenario, as depicted in figure \ref{fig:topologyC}), two systems are running in a subnet together with one attacker.

        Node 21 is accessible only by the credentials \textit{alice:starwars}. These credentials are well-hashed on node 21, but like in reality, user credentials can be reused throughout multiple devices and locations. In this topology, the password of the user Alice is also stored on node 22, but in this case, the password is just stored with a weak hash, which means it can be brute-forced quickly.
        Node 21 also runs the same vulnerable service as node 3 in Scenario A. After exploiting the vulnerability, an attacker can gain read access to the \textit{shadow} file. In Unix systems, the shadow stores the hashes of all the system's user passwords.
        A \textit{\$1\$} leading the hash is indicating an MD5 hash. If the user uses a weak password, together with this weak hash, the attacker can use a tool like John the Ripper \cite{SolarDesignerOpenWall.} to crack the password. After the attacker gets access to these new credentials, he can use them to log into the other supposedly secure system. Ultimately, it is possible to compromise both systems using exploits and to crack found password hashes.

    \subsection{Penbox}
\begin{figure}[htbp]
\centerline{\includegraphics[width=\columnwidth]{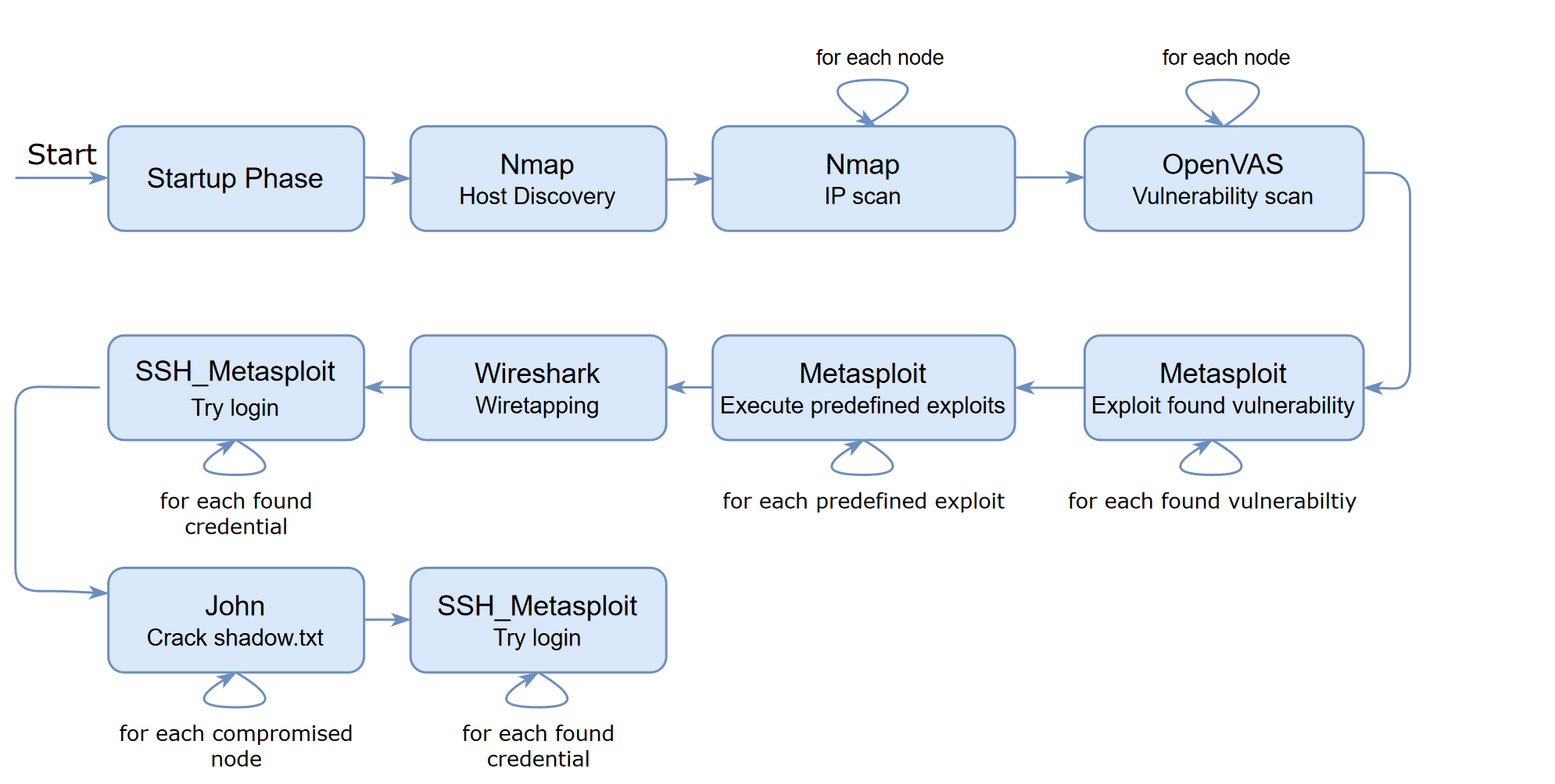}}
\caption{Penbox execution chain}
\Description[ToDo]{ToDo}
\label{fig:exec_chain}
\end{figure}

    Penbox needs a specification of an attack tree. To apply Penbox successfully to the scenarios, a decision tree is defined. For readability reasons, the decision tree is presented as an execution chain, as can be seen in Figure \ref{fig:exec_chain}. Before the next phase is selected the action is performed on every reachable and discovered node in the network.

    \subsection{NASim and applied modifications}
        An environment for the agent to interact with is needed to perform reinforcement learning. Real networks are unsuitable due to their slow interaction speeds. The delays in real networks, combined with the longer execution times for specific actions, would significantly slow down the agents' training process, which demands extensive training. To simulate the scenarios, NASim is used. NASim was one of the early environments to apply reinforcement learning to penetration testing \cite{Schwartz.2019}. NASim's simulation is very far-reaching and tries to cover the most basic penetration testing mechanics. Internally, NASim represents the network and all its hosts by a single matrix, including information like which service is running on each machine. Still, in its current state, it cannot simulate the scenarios of Exploits, Wiretapping and Post-exploitation because these are too specific. This is why for this work NASim is extended by the following:
        \begin{itemize}
            \item \textit{Exploits key-and-lock principle}\\ In its current version, when executing an exploit on an existing service, the exploit has a probability of ending with success. This requires making an assumption about and quantifying the chance of a service being exploitable. In this case, an RL agent would most likely just learn these probabilities. This poses the problem that it is not trivial to obtain real-world probabilities. One might use internet scanners such as Shodan \cite{Schwartz.2019}, to obtain the spread of services. Still, there is a strong bias towards internet-facing services such as ssh and http, and thereis no way to estimate the likeyhood for a service to be vulnerable. 
            Still, in most cases, an exploit works with the key-and-lock principle. Each exploit works for a specific service with a specific version. The previous NASim implementation leads to the following: In the simulation, it is possible to repeat the same exploit multiple times, and if performed multiple times, the attacker gains access to the machine due to the probability. This does not represent reality: if the service is vulnerable, the exploit works, otherwise not. The probability would only make sense to generate the topology and set 10 out of 100 FTP services to be exploitable in the topology. To solve this problem, the authors made it possible to set the probability of an exploit to only 0 or 1. 
            Additionally, instead of naming the service by its protocol, such as FTP, a specific version can be specified, like \textit{proftpd-5} and \textit{proftpd-1.3.3d}, where \textit{proftpd-1.3.3d} is only exploitable with the suitable exploit. 
            \item \textit{Vulnerabilities}\\ The concept of vulnerabilities was also added. NASim does not provide any vulnerability scan. The vulnerabilities are represented in the Hostvector by a boolean,
            and need to be set in the configuration of the network. The scan behaves equally to the already existing OS scan. The information about the vulnerability is not necessary to execute the exploit successfully. Yet, vulnerabilities must exist on the machine for an exploit, which requires the vulnerability to work. \\
            
            \item \textit{Credentials and Wiretapping}\\ Credentials and wiretapping were also added. Both are not present in the original NASim version but are necessary to simulate the scenarios of wiretapping and post-exploitation. Wiretapping is an action which can return user credentials. Wiretapping can be performed on a target on its own. In the best case, it will return a predefined amount of credentials. These credentials are defined inside the  configuration file. Found credentials will be stored inside the host-vector, the internal representation of host information in NASim. Since the attacker should always find the same credentials in one subnet with wiretapping, it does not matter on which target he performs the action, inside the subnet.\\
            To represent credential pairs internally, IDs from 1 to 9 were used. Each number presents one credential pair. If there are multiple credentials present, the IDs are concatenated as unordered tuple. For example, if there are two credentials, e.g. \textit{user1:pass1} and \textit{user2:pass2}, they are represented in the simulation by \textit{12} or \textit{21}, which is equal.\\ 
        \end{itemize}
        In the standard implementation of NASim, the attacker starts from outside the network. The presented scenarios simulate an insider attack, so an initial host is added  to the subnet. This host is then used as an entry point to the network and has its own initial exploit.   
        With these changes, NASim can simulate all three scenarios.

    \subsection{Reinforcement Learning Algorithms}
        \subsubsection{Q-learning}
        Q-learning is an action-value algorithm with a tabular representation, which means it tries to map a value for each possible action. Based on this value, the best action is chosen. To be more specific, Q-learning is a temporal-difference off-policy RL algorithm. The action-value function $q(s_t,a_t)$ starts with initial values, which are then updated over the training. Off-policy means the agent learns the best policy $\pi^*$ while exploring the environment using another policy $\pi^b$. \\
        It is not necessary to use a tabular representation. Instead, an approximated representation can be used. This, for example, can be a neural network and is done in Deep Q-learning or short DQN. \\
        With the given formula of reinforcement learning, the agent can gradually construct an approximation of the true action-value function $q(s_t,a_t)$ by gradually updating its estimation by the following so-called, Q-function: 
        $$ Q(s_t,a_t) \leftarrow Q(s_t,q_t)+\alpha[r_t + \gamma \textit{max}_x Q(s_{t+1},x ) -Q (s_t,a_t)]$$
        Where:
        \begin{itemize}
            \item $\alpha \in \mathbb{R}$  is the learning rate or step size and defines how fast the agent will learn. If $\alpha = 0$, the agent learns nothing from newly performed actions. If $\alpha = 1$, prior learned knowledge is ignored, and the most recently performed action is essential.
            \item $\gamma$ is the discount factor, giving more or less weight to future rewards.
        \end{itemize}
        With standard Q-learning, the problem of exploitation and exploration is still present. To solve this, Q-learning with $\epsilon$-greedy can be used. This method tries to bring randomness into the process of selecting actions. This way, the known actions are not exclusively performed. The algorithm usually prefers known actions as they have already yielded rewards. Some new actions need to be tried to include them in the known actions by gaining information on their performance. This helps the algorithm not to get stuck at a local optimum. $\epsilon = 1$ would represent random actions every time, while $\epsilon = 0$ means no random actions at all. Decreasing the $\epsilon$ over time is standard, as it usually leads to better results. So, in the beginning, the algorithm uses a high $\epsilon$ and, at the end, very small ones, i.e. first explore more, then focus on the known actions and their known yields.
        An already provided Q-learning algorithm of NASim is used \cite{Schwartz.2019}. 
        


        \subsubsection{DQN}
        DQN stands for Deep Q-learning Network and is very similar to the Q-learning. It was first presented by Mnih et al. \cite{Mnih.2015}. The main difference to Q-learning is that it uses a single Deep Neural Network to estimate the action value. DQN approximates the state-action value function such that $Q(s,a;\theta)\approx q_\pi(s,a)$, where $\theta$ denotes the neural network's weights. The neural network (short NN) uses a state $s_t$ and outputs $|A|$ scalars corresponding to the state-action values of $s_t$ \cite{Farebrother.9292018}. DQN then tries to optimize the NN. 
        
        DQN includes all of the Q-learning Hyperparameters but, additionally, there are the parameters of the Neural Network: 
        \begin{enumerate}
            \item \textbf{optimizer} that adjusts the network's parameters during training to minimize the loss function, helping the network learn and improve its performance on a specific task.
            \item \textbf{batch size} defines the number of samples propagating through the network.
            \item \textbf{replay size} is the maximum number of experiences stored in the replay memory.
            \item \textbf{hidden size} is the number of hidden layers of the Neural Network.
            \item \textbf{target update frequency and update interval} is how often the NN weights are updated. Adjusting this can avoid runaway feedback. 
        \end{enumerate}
        The DQN version implementation by Chainerrl \cite{YasuhiroFujita.2021} and the DQN algorithm of NASim \cite{Schwartz.2019} were used for the experiments.

        \subsubsection{A3C}
        Besides Q-learning, Asynchronous Advantage Actor Critic (A3C) is a newer and more advanced approach. It was developed by Google’s DeepMind \cite{Mnih.2016}. 
        \begin{itemize}
            \item \textbf{Asynchronous} because the algorithm is not trained on a single environment simultaneously. A3C trains multiple independent workers on different environments at the same time. Each worker shares the knowledge over a global network, which is asynchronously updated after a predefined amount of steps $t_{max}$. 
            \item \textbf{Advantage} means instead of the policy gradient, with its discount factor $gamma$, A3C uses the Advantage value to determine which action was most rewarding.
            \item \textbf{Actor-Critic} combines the advantages of value-iteration and policy-gradient methods. The algorithm maintains a value function $V(s,\theta)$ to update the optimal policy $\pi(a_t|s_t;\theta)$, where $\theta$ are the parameters of the policy. As the authors describe, the performed update can be seen as \\ $\nabla_{\theta'} log \pi(a_t | s_t; \theta')A(s_t,a_t;\theta,\theta_v)$ where $A(s_t,a_t; \theta,\theta_v)$, is an estimate of the advantage function given by $\sum_{i=0}^{k-1} \gamma^i r_{t+i}+\gamma^k V(S_{t+k}; \theta_v) - V(s_t;\theta_v)$ and k can vary but has the upper bound of $t_{max}$. \cite{Mnih.2016}
        \end{itemize}
        Other than most reinforcement algorithms using neural networks, the training of A3C is only done on the CPU and not on a GPU. Still, A3C is a fast and robust algorithm with promising results. 
        The A3C version implementation Chainerrl \cite{YasuhiroFujita.2021} was used for the experiments. If the hyperparameters are provided, they are given in the following order: model, learning rate, final epsilon, reward\_scale factor, beta, gamma, and alpha.
        

        \subsubsection{Reward Model}
        The reward model has a significant impact on the success of an agent. If the reward model is poorly defined, the agent can find ways to maximise its reward without behaving as intended, as can be seen in the example of the race game CoastRunners. The intended goal was to win the race, but the agent collected items instead of finishing them because it was more rewarding. \cite{ClarJack.2016}. 
        The goal of the scenarios is to earn root privileges. The agent achieves a positive reward if and only if a machine is completely compromised. All actions have a small cost, represented by a negative reward, as shown in table \ref{tab:reward}. As mentioned before, an entry point host is used, and its compromise is rewarded with a reward of 50. Resulting in the following maximal positive obtainable reward per scenario:
        \begin{itemize}
            \item Scenario A: 350 
            \item Scenario B: 250
            \item Scenario C: 150 
        \end{itemize}
        
\begin{table}
\centering
\caption{\label{tab:reward} Reward Model}
\begin{tabular}{l|l}
 & reward \\ \hline
\multicolumn{1}{l|}{exploit action} & -3 \\ 
\multicolumn{1}{l|}{post\_exploit action} & -3 \\ 
\multicolumn{1}{l|}{initial phase exploit} & -1 \\ 
\multicolumn{1}{l|}{all scans} & -1 \\
\multicolumn{1}{l|}{wiretapping} & -1 \\
\multicolumn{1}{l|}{compromised host} & 100 \\
\end{tabular}
\end{table}
    \subsection{Permutations}
        First, small tests have shown that agents overfit when trained on one scenario. They learn their strategies by heart. The agents will skip the scanning stage to maximize their reward. This makes no sense in reality for penetration testing agents. To address this, instead of training on just one scenario, permutations of each scenario are created and rotated while training. After 100 steps are done or after an episode is successfully solved, a new permutation is selected.\\
        To create the permutation, the host order is switched inside the simulation. In a real network, it would be equivalent to a change of IP address. Additionally, each scenario is filled up to a fixed host size of 5 hosts, one attacker, and four other hosts. Both scenarios that do not have four hosts will be filled with empty hosts, which have no services. This has the positive effect of having a discrete state and action space. Reinforcement learning problems with discrete action spaces are easier and better to solve. The same applies to the state space.\\
        The attacker and his entry point is not involved in the rotation. This results in $4! = 24 $ different permutations of each scenario, one permutation is defined as environments. Overall, a set of $3\cdot24=72$ environments is created. An important note here is that topologies B and C theoretically include 12 different environments because they have two empty hosts, which results in the same network topology.
        IP switching might sound like it does not significantly alter the scenario. This is not the case for NASim and the RL agents. By changing the IP, the state of the simulation completely changes, and an agent who is just trained on one permutation cannot solve the other one.

    \subsection{Grid Hyperparameter Search}
        A large grid hyperparameter search evaluates and compares the different RL algorithms with different stages of complexity. The first stage is to train and solve all 24 permutations of one scenario. The second stage is to solve all permutations of all 72 permutations of every scenario. The third stage split the set of permutations, the scenarios separated, into a test and train set. This will show how well the agent can generalize.
        
        Before performing such large and heavy computational training, smaller tests were performed to ensure that the agents could handle multiple switching environments. The initial smaller tests include slowly increasing the number of permutations and training steps. Q-learning and A3C pass these initial tests, so for each of them, 1296 different hyperparameter combinations were selected, see appendix Table \ref{tab:641}, and each agent is trained for 8.000.000 training steps. The training was performed on a CPU cluster. A single agent of A3C needs 80 min to train, with 26 CPU kernels of an AMD EPYC 7742 (64-Core Processor).  $1296 \cdot 80 min $ would be around 70 days to train all agents on one Scenario with 24 environments. Using up to 1000 CPU kernels once at a time, the authors could train the agents in a feasible amount of time for every Scenario. 

        \begin{table*}[!t]
        \centering
        \caption{\label{tab:641} A3C and Q-learning Hyperparameter  }
        \makebox[1 \textwidth][c]{       
        \resizebox{1 \textwidth}{!}{
        \begin{tabular}{l|l|l}
        & \bfseries A3C &\bfseries Q-learning \\ 
        \hline
        \textbf{Model} & FFSoftmax, FFMellowmax & -   \\ 
        \textbf{Learning Rate}  & 0.3, 0.1, 0.01, 0.0001  & 0.8 , 0.5, 0.3, 0.1, 0.01, 0.001 \\
        \textbf{Final epsilon} & 0.3, 0.1, 0.01 & 0.8 , 0.5, 0.3, 0.1, 0.01, 0.001 \\
        \textbf{Reward\_scale factor} & 1, 0.01 & - \\
        \textbf{Exploration Steps} & - & 80.000, 800.000, 2.000.000, 4.000.000, 6.000.000, 8.000.000 \\
        \textbf{Beta} & 0.1, 0.01, 0.001 &- \\
        \textbf{Gamma} & 0.5, 0.9, 0.99 & 0.01, 0.4, 0.7, 0.9, 0.99, 0.999 \\
        \textbf{Alpha} & 0.99, 0.9, 0.8 & - \\ 
        \end{tabular}
        }
        }
        \end{table*}
        
        The initial tests also show that DQN starts failing when trained on two permutations. It cannot even solve one of the two environments when being trained on both, so it was left out of the Large Hyperparameter search and got its own smaller Hyperparameter search. Instead of being trained on 24 permutations, it is trained on just two with the hyperparameter in Table \ref{tab:777} (appendix) for $100.000$ steps. Later, the best 16 agents out of the 384 will be trained for $1.000.000$ training steps. To ensure the training size is large enough. 

                \begin{table*}[!t]
        \centering
        \caption{\label{tab:777} DQN Hyperparameter }
        \makebox[1 \textwidth][c]{       
        \resizebox{1 \textwidth}{!}{
        \begin{tabular}{l|l}
        & \bfseries DQN \\ 
        \hline
        \textbf{Qfunc} & [”FCStateQFunctionWithDiscreteAction”,2,50] [”FCLSTMStateQFunction”, 2, 50] \\
        \textbf{Optimizer}  & [adam,1e-2], [adam,1e-4] \\
        \textbf{Final epsilon} & [”LinearDecayEpsilonGreedy”,0.1,9*10**4],[”LinearDecayEpsilonGreedy”, 0.1, 6*10**4] \\
        \textbf{Replay buffer} & 1.000, 100.000 \\
        \textbf{Gamma} & 0.7, 0.9, 0.99\\
        \textbf{Target update interval} & 100, 1000 \\
        \textbf{Update interval} & 1, 100 \\
        \textbf{Replay start size } & 500, 1000 \\ 
        \end{tabular}
        }
        }
        \end{table*}
        
\section{Results}        
        \subsection{DQN}
\begin{figure}[ht]
        \centerline{\includegraphics[scale=0.6]{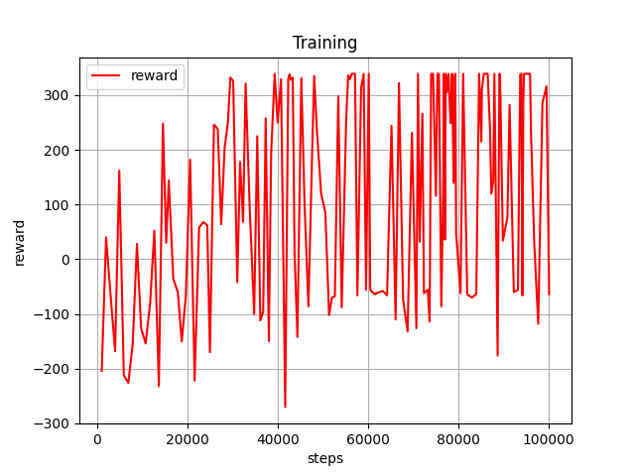}}
        \caption{DQN agent trained on one permutation of Scenario A, with the following Hyperparameter:  q\_func: ['FCStateQFunctionWithDiscreteAction', 2, 50] optimizer: ['adam', 0.0001] replay\_buffer: 1000 gamma: 0.7 explorer ['LinearDecayEpsilonGreedy', 0.1, 60000] target\_update\_intervall: 1000 update\_intervall: 1 replay\_start\_size: 1000.}
        \label{fig:DQN_tests_1env_10}
        \Description[ToDo]{ToDo}
        \end{figure}

        The first impression about DQN is due to its neural network action selection; it should be able to solve multiple environments and archive some generalisation. The experiments of this work proved different. The results were so underwhelming that even multiple implementations were tested to ensure it was not an implementation issue from the authors (Chainer \cite{Fujita.2021} and NASim \cite{Schwartz.2019}). As seen in Figure \ref{fig:DQN_tests_1env_10}, DQN can solve a single environment (in this case, Scenario A) when trained on it.

\begin{figure*}[ht]
        \centering
        \setkeys{Gin}{width=0.2\linewidth}
        \subfloat[]{\includegraphics{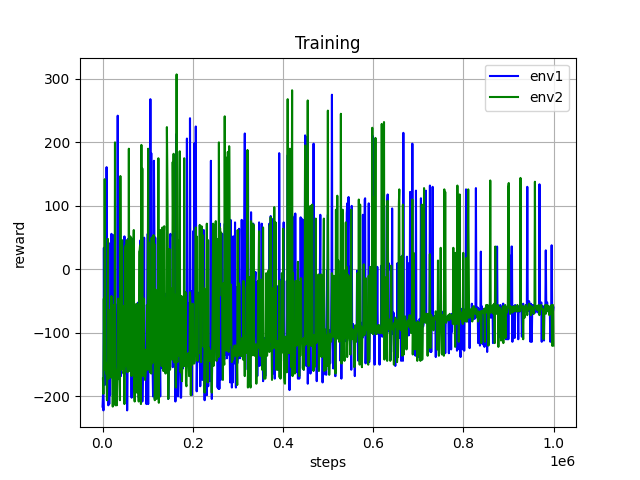}} \label{test}
        \hfill
        \subfloat[]{\includegraphics{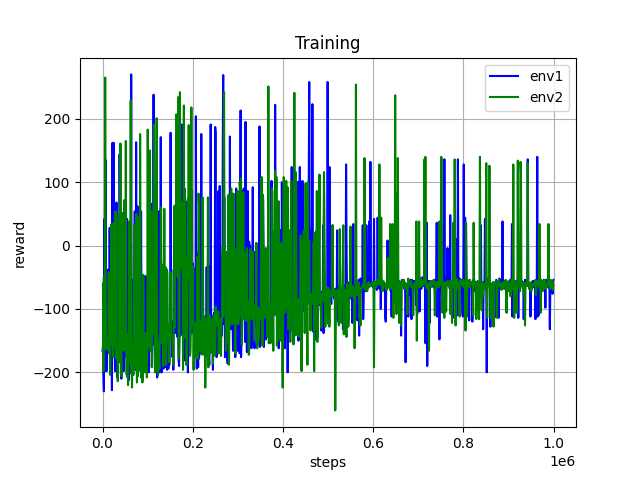}}
        \hfill
        \subfloat[]{\includegraphics{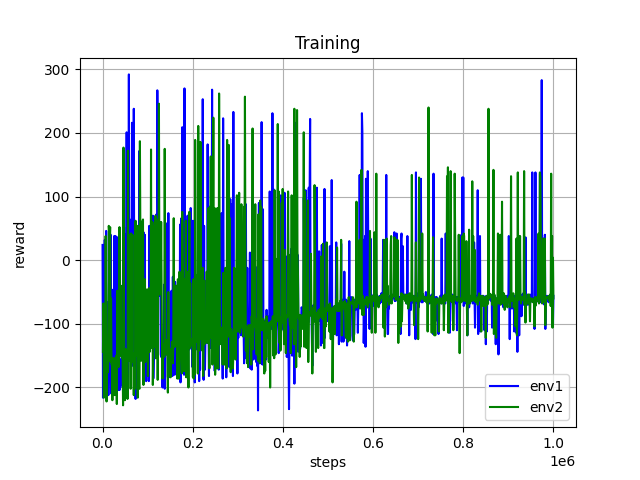}}
        \hfill
        \subfloat[]{\includegraphics{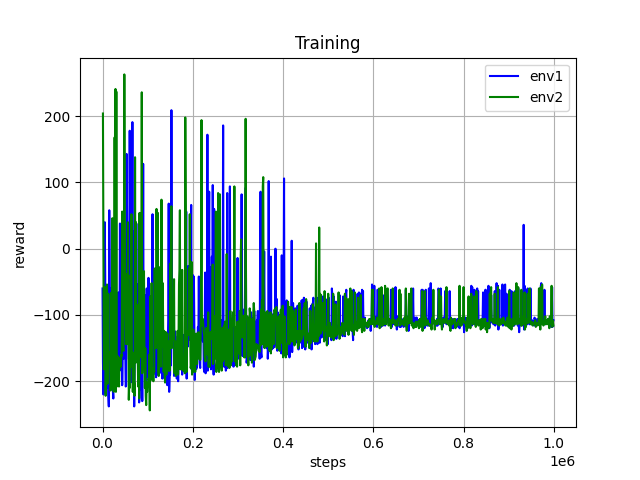}}
        \hfill
        \caption{Some randomly picked training graphs of the best 16 DQN agents with 100.000 training steps, now trained for 1.000.000. Showing that DQN is not able to solve multiple permuations of one Scenario.}
        \label{fig:DQN_tests_3}
        \Description[<short description>]{<long description>}
        \end{figure*}

        As soon as the environment amount is increased from one to two, none of the 384 trained agents produces an acceptable positive reward, even if the training size is increased by a factor of 10. During training, the reward continues to deteriorate until it settles around -100 almost every time, regardless of the agent hyperparameters, see figure \ref{fig:DQN_tests_3}. This is true for all 384 agents, whether trained for $10.000$ or $100.000$ training steps. DQN can not solve stage one of the three stages in the author's experiments. It is not considered for the other stages. 


        \subsection{Stage 1: 24 permutations of one scenario}
\begin{figure*}[ht]
                \centering
                \setkeys{Gin}{width=0.27\linewidth}
                \subfloat[Scenario A; Q-learning]{\includegraphics{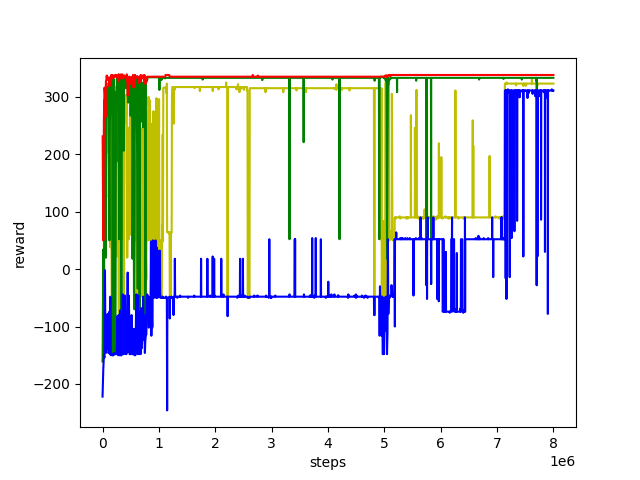}}
                \hfill
                \subfloat[Scenario B; Q-learning]{\includegraphics{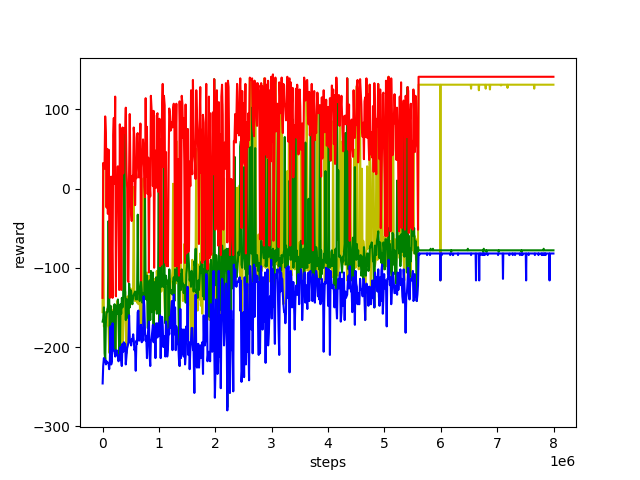}}
                \hfill
                \subfloat[Scenario C; Q-learning]{\includegraphics{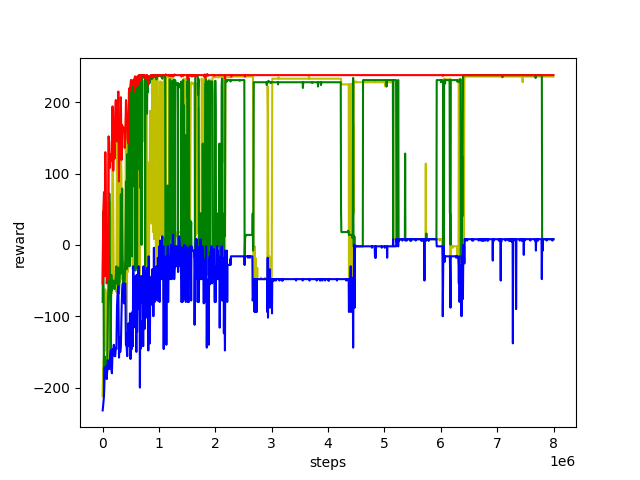}}
                \hfill
                \subfloat[Scenario A; A3C]{\includegraphics{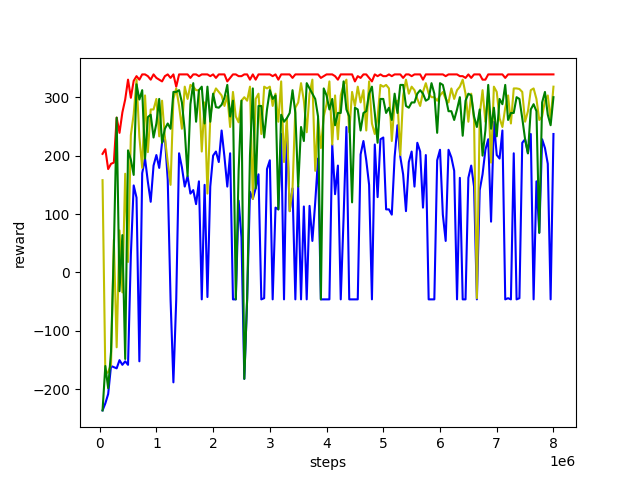}}
                \hfill
                \subfloat[Scenario B; A3C]{\includegraphics{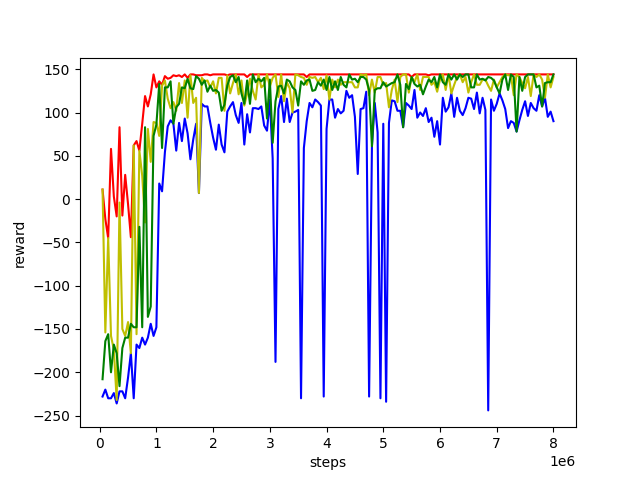}}
                \hfill
                \subfloat[Scenario C; A3C]{\includegraphics{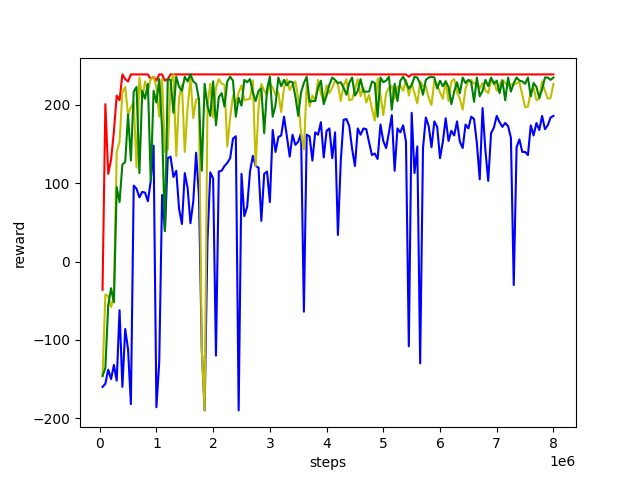}}
                \caption{Training of best-performing agents per topology. Hyperparameter can be seen in Table \ref{tab:666} (appendix). Maximum Red; Minimum Blue; Permutation 4 Yellow; Permutation 15 green.}
                \label{fig:images3}
                \Description[ToDo]{ToDo}
            \end{figure*}

            Each agent was trained on all 24 permutations of one scenario simultaneously, which means for Q-learning, after a successfully solved episode, the maximum number of hosts are compromised, or after 100 steps, the permutation is changed in a fixed order. For A3C, an agent was trained on each permutation in parallel. 
            The training of the best-performing agents per topology can be seen in Figure \ref{fig:images3}. Red shows the maximum reached reward of all permutations, while blue shows the minimum. Permutation 4, yellow and permutation 15, green, were selected to demonstrate concrete environments and their reward due to the training. The most important value is the minimum. It indicates if all permutations can be solved with an optimal attacking path. An optimal attacking path is close to the maximum possible reward. For Q-learning in Figure \ref{fig:images3} (a), (b) and (c) can be seen that Q-learning finds perfect attack paths (red), but it is not able to solve every permutation, except for Topology A. On Topology A, Q-learning can solve 24 out of 24 permutations after the training is finished. Furthermore, 24 permutations are likely solved after 7000000 training steps. In all three cases, the minimum archived reward improves over the training. The average of all permutations Q-learning can solve is higher than the average reward and amount of actions needed by Penbox, see Table \ref{tab:232}.
            A3C, on the other hand, surpasses Q-learning. In Figure \ref{fig:images3} (d), (e) and (f), the best agents of A3C can be seen. In contrast to Q-learning, the minimum always increases and is getting way closer to the maximum possible reward than Q-learning. The best resulting agents of A3C can solve all 24 permutations of one topology. It can be seen that A3C trained on single scenarios exceeds even the results of Q-learning compared to Penbox, see Table \ref{tab:232}.

        \begin{table*}[!t]
        \caption{\label{tab:666}  Best performing Agents of each Scenario (Stage 1)}
        \makebox[1 \textwidth][c]{       
        \resizebox{1 \textwidth}{!}{
        \begin{tabular}[]{ll|l|l|l|l|l|l|l|l}
        \textbf{Q-learning} & &\bfseries Model &\bfseries Learning Rate  &\bfseries Final epsilon &\bfseries Reward scale factor &\bfseries Exploration Steps &\bfseries Beta &\bfseries Gamma &\bfseries Alpha \\ 
        \hline
        & \textbf{Scenario A} & - & 0.01 & 0.001 & - & 80000 & - & 0.1 & -  \\ 
        & \textbf{Scenario B} & - & 0.1 & 0.001 & - & 2000000 & - & 0.999 & -  \\ 
        & \textbf{Scenario C} & - & 0.01 & 0.001 & - & 80000 & - & 0.1 & -  \\
        \textbf{A3C} &    &&&&&&&        \\ \hline
        & \textbf{Scenario A} & FFSoftmax & 0.0001 & 0.01 & 0.01 & - & 0.01 & 0.5 & 0.99  \\
        & \textbf{Scenario B} & FFMellowmax & 0.0001 & 0.01 & 0.01 & - & 0.001 & 0.99 & 0.9  \\
        & \textbf{Scenario C} & FFMellowmax& 0.01 & 0.3 & 0.01 & - & 0.01 & 0.99 & 0.9  \\
        \end{tabular}
        }
        }
        \end{table*}
        
        \subsection{Stage 2: 72 permutations of all three scenarios}
        Q-learning cannot solve all permutations of every scenario, so only A3C is selected for the second stage. The next step is to solve all scenarios and their permutations with just one agent. The agents are trained on all 72 permutations instead of just 24. None of the 1296 A3C agents gets close to the results from the previous stage. Each agent can solve some permutations very well but fails to solve others, such as Q-learning being successful in Scenario A but failing some permutations in Scenario B. The best agent with \textit{FFSoftmax, 0.0001, 0.01, 0.01, 0.001, 0.99, 0.99} reaches an average reward of 226.25 on Scenario A, 136 on Scenario B and 125 on Scenario C. When compared with Table \ref{tab:232}, from the previous stage, this is a significant decrease.\\
        More environments also require larger training. An additional training was performed, in which the training steps were increased to 200,000,000 instead of 8,000,000. Because such training is considerably more time intensive, just the 12 best agents from the previous training were selected. When increasing the training steps, the A3C agent with the following hyperparameter: \textit{FFSoftmax, 0.0001, 0.01, 0.01, 0.001, 0.99, 0.99} can solve all permutations of each scenario, with an optimal attack path. The agent's result can again be seen in Table \ref{tab:232}.
        
        \subsection{Stage 3: Generalization of A3C on one scenario}
        Lastly, the generalization is tested. The best agent for each scenario from stage 1 is selected and trained. Instead of all 24 permutations, the permutations get split into a train and test set. To see how many train environments are required to solve all 24 permutations. Multiple trainings were performed, starting with 12 permutations inside the train and test set. If the resulting agent cannot solve all permutations, the training is repeated, and one environment from the test set is moved to the training set. The results are the following:
        \begin{itemize}
            \item On Scenario A, the agent needs to be trained on 20 to solve 24 permutations
            \item On Scenario B, the agent  needs to be trained on 12 or more to solve 24 permutations
            \item On Scenario C, the agent  needs to be trained on 18 or more to solve 24 permutations
        \end{itemize}

        \begin{figure*}[ht]
                \centering
                \setkeys{Gin}{width=0.23\linewidth}
                \subfloat[Scenario A, test-set size of 4]{\includegraphics{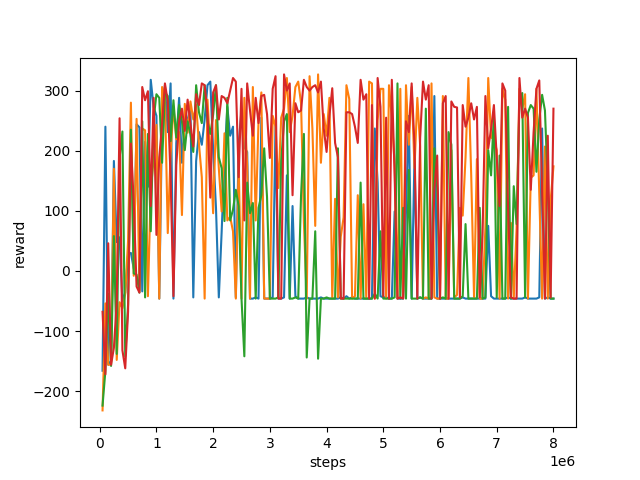}} \label{test}
                \hfill
                \subfloat[Scenario A, test-set size of 2]{\includegraphics{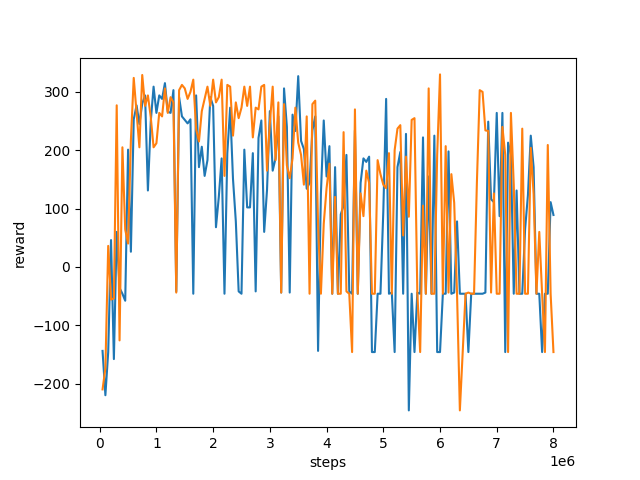}}
                \hfill
                \subfloat[Scenario B, test-set size of 12]{\includegraphics{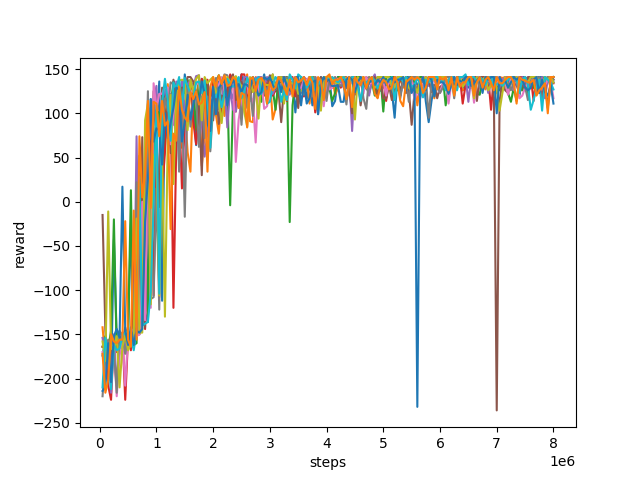}}
                \hfill
                \subfloat[Scenario C, test-set size of 6]{\includegraphics{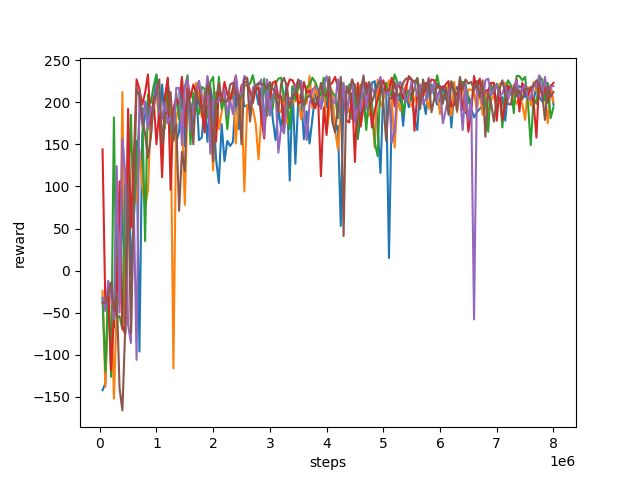}}
                \hfill
                \caption{ On Scenario C, the agent  needs to be trained on 18 or less to solve 24 permutationse training time for the different test set permutations to estimate the agent's generalization capability for each scenario. Each color represents a different permutation.}
                \label{fig:images6}
                \Description[<short description>]{<long description>}
        \end{figure*}

        In Figure \ref{fig:images6}, it can be seen how the reward of the training set behaves. Each colour represents one permutation in the test set. In Scenario A, Figure \ref{fig:images6} (a) and (b), the training looks noisy. At one point in the training, the agent can solve the permutation successfully and starts to fail after further training. Still, the final resulting agent can solve all permutations after the training. While in (c) and (d), the unseen cases stay constant after a specific threshold. Very interesting to see is that in scenario A, trained with 22 permutations, a clear overfitting can be seen after 1,000,000 training steps.

\begin{table*}
\centering
\caption{\label{tab:232} Penbox compared against Q-learning and A3C. For Q-learning, just the average of correctly solved permutations is shown.}
\makebox[1 \textwidth][c]{       
\resizebox{1 \textwidth}{!}{
\begin{tabular}{ll|l|l|l|l}
Scenario A & & Penbox & Q-learning  & A3C single scenario & A3C all scenario (72 permutations) \\ \hline
   & solve        &  24/24  & 24/24   & 24/24   &  24/24    \\
   & avg reward  (max 334)&317.5  & 327.08  & 301.875 &  304.42   \\
   & avg actions (min 9) &17.5   & 12.42   & 17.375  &  22.333   \\
Scenario B &             &        & & &            \\ \hline
   & solve       & 24/24  & 14/24    & 24/24 &  24/24  \\
   & avg reward  (max 140) &107.5   & 135.86   &  136.75 & 129.71 \\
   & avg actions (min 6)& 21.5   & 11   & 6.5 &  11.625 \\
Scenario C &             &        &   & &          \\ \hline
   & solve        &24/24 & 16/24     & 24/24   &   24/24 \\
   & avg reward   (max 236)& 201.5  & 232.5   & 223.29 & 208.5  \\
   & avg actions (min 6) & 23.5   & 10.5     & 13.46 &  21.5
\end{tabular}
}
}
\end{table*}

\section{Discussion}
The experiments show that the A3C agent can solve all permutations and outperform Penbox, a standard automated penetration testing tool, regarding the number of actions needed. The performed actions are very close, if not perfect, to the optimal attack path. Additionally, the A3C agent was able to generalize and even solve unseen permutations. It is important to not just consider successful solves as an evaluation metric, but to also keep track of the number of actions needed to solve the scenario. If the agent can solve an environment, but requires more actions than a random agent or a predefined decision tree, the RL-based approach offers no advantage. In a real-world setting the amount of actions is relevant, as an increased amount of actions results in a longer testing time and a louder and therefore easier detectable attack, which decreases the chance of success. \\

With Penbox, a potential interface is provided to execute the actions in real networks, and the gathered information can be passed back into the simulated NASim environment.
However, when applying the agents to real-world environments, the following points need to be considered: (1) The agents are trained in a perfect world, without connectivity problems, crashes, timeouts or defence mechanisms. (2) The topologies are similar, and their action and state space is fixed. (3) The agents are limited to the methods learned in the training.\\ 

In reality, an attacker might face connectivity problems, crashes or timeouts. These issues should be handled by a wrapper interface and not by the RL agents to reduce the action and state space. Additionally, the experiments don't provide any active defences or countermechanisms. This makes it much easier for an attacker to reach their goals. In the real world, many networks provide intrusion detection systems and firewalls or even honeypots and blue teams. The authors were not emphasising on the stealthiness of the agents, so there are no counter defence actions applied.\\

The discrete action and state space limit the agent's capability of interacting with networks. The agent can handle networks with a size of exactly the amount of hosts it was trained for and a fixed amount of actions. To address this, the agent needs to be trained using a continuous action and state space, which can negatively impact the agent's performance. Aside from that, the discrete action and state space can be expanded to common network subnet sizes, like 256. This would result in an explosion of the possible actions and states, thus greatly impacting the performance.
Considering Scenario A, an exploit service mapping could be enough to make reinforcement learning superfluous. The mapping of exploits to vulnerable services could be provided and not be necessary to be learned and performed by the agent. This will speed up the training, increase performance and allow the agent to focus on other tasks. Although the provided tasks require rather trivial actions to be solved, showing that these dependencies could be found by the agents is a necessary requirement for the success in scenarios with more complex dependencies. \\

The fact that the agents can only solve similar topologies and are limited to their trained attacking methods has to be considered when applying the agent to real networks. Similar to human testers, the RL algorithms are limited to a specific skill set, which they can improve by new learning. 
While it would be interesting to measure how the agents perform in other scenarios than they were trained on, in this work each scenario requires different exploitation methods, leading to different sets of actions required. Therefore, the agents would not perform well in the other scenarios without retraining. These problems of retraining and knowledge transfer are broad topics and may be the scope for feature research. 
These scenarios in this work only represent a small set of penetration testing techniques, and the resulting agents are limited to the actions from the training, requiring retraining to extend an agent with new techniques.
The amount of 72 environments used in this work are not sufficient to achieve greater generalization. This was already indicated by the overfitting in Stage 3. To achieve a more generally applicable agent, a large number of different topologies which differ in size and complexity is required.\\

The reward function used in the experiment rewards the successful compromise of a host. It would be reasonable to reward information gathering, because information about the target can be valuable for the attack, but this was not the scope of this work. \\
Simply changing the IP address of a node does not seem to alter the scenario significantly. However, this work shows that most agents fail with such an easy task, as seen for DQN. Regarding the DQN agents, the authors consider that the permutation of environments reverts the respective neural network optimisations. \\


\section{Conclusion}
In this paper, the authors have developed three different penetration testing scenarios. To solve these scenarios with reinforcement learning, the Network Attack Simulator, NASim, was modified to represent these scenarios inside a simulation. A3C, Q-learning and DQN were used as algorithms compared against an already existing automated penetration testing method using decision trees. An extensive hyperparameter grid search was performed on two algorithms to find the best agents. The experiments show that A3C can solve all scenarios and find the optimal attack paths.
In most cases, A3C was able to outperform the baseline algorithms with respect to the amount of actions. It was shown that A3C agents are able to achieve a good generalization, while Q-learning and DQN were less successful when being trained on multiple environments. Nevertheless, for the A3C agents a limited state and action space was utilized and the interaction was done on rather small scenarios. A tendency to overfitting was observed, which implies that a larger variety of training scenarios is needed. Therefore, a scenario generator might address this problem in future work. Overall, the results are promising for the future of autonomous penetration testing, and while a lot of research is still due, it could be shown that this approach is promising.

\section{Acknowledgments}
This research was supported by the German Federal Ministry of Education and Research (BMBF) through the Open6GHub project (Grant 16KISK003K).

\bibliographystyle{IEEEtran}
\bibliography{references.bib}
\end{document}